\documentclass[pdftex,twocolumn,epjc3]{svjour3}          

\RequirePackage[T1]{fontenc}

\smartqed  

\RequirePackage{graphicx}
\RequirePackage{mathptmx}      
\usepackage{amsmath}
\RequirePackage{flushend}
\RequirePackage[numbers,sort&compress]{natbib}
\RequirePackage[colorlinks,citecolor=blue,urlcolor=blue,linkcolor=blue]{hyperref}
\usepackage{orcidlink}
\usepackage{physics}
\usepackage{float}

\journalname{Eur. Phys. J. C}

\begin{document}

\title{Insights into Neutron Stars from Gravitational Redshifts and Universal Relations}


\author{Sagnik Chatterjee \orcidlink{0000-0001-6367-7017}\thanksref{addr1}
        \and
        Kamal Krishna Nath \orcidlink{0000-0002-4657-8794}\thanksref{addr2}
}

\thankstext{e1,addr1}{e-mail: sagnik18@iiserb.ac.in}
\thankstext{e2,addr2}{e-mail: kknath@niser.ac.in}

\institute{Indian Institute of Science Education and Research Bhopal, Bhopal 462066, India \label{addr1}
\and
School of Physical Sciences, National Institute of Science Education and Research, An OCC of Homi Bhabha National Institute, Jatni-752050, India \label{addr2}
}

\date{Received: date / Accepted: date}

\maketitle

\begin{abstract}
The universal relations in neutron stars form an essential entity to understand their properties. The moment of inertia, dimensionless tidal deformability, mass quadrupole moment, and oscillation modes are some of the properties that have been studied previously in the context of universal relations. All of these quantities are measurable; thus, analyzing them is of utmost importance. In this article we provide new universal relations in the context of a neutron star's gravitational redshift. Using the redshift measurements of RBS 1223, RX J0720.4-3125, and RX J1856.5-3754, we provide theoretical estimates of moment of inertia, dimensionless tidal deformability, mass quadrupole moment, the mass of the star times the ratio of angular frequency over the spin angular moment, and the average of the speed of sound squared. In the case of the redshift measurement of RX J0720.4-3125, we found that the theoretical estimate using universal relations aligns closely with the Bayesian estimate. Our findings indicate that such theoretical predictions are highly reliable for observations with low uncertainty and can be used as an alternative for statistical analysis. Additionally, we report a violation of the universality of the dimensionless tidal deformability and average of the speed of sound squared with respect to the gravitational redshift. Our calculations further indicate that, under current astrophysical constraints, the maximum gravitational redshift attainable by neutron stars does not exceed $0.763$.

\end{abstract}

\section{Introduction}\label{sec:intro}
Neutron Stars (NSs) are compact objects having central densities up to 2-8 times that of the nuclear saturation density ($n_s$). Such intermediate density ranges are yet to be probed by terrestrial laboratories, making the core of the NS a natural laboratory to study such densities. The quest to study the core of NSs is often accompanied by constraining equations of state (EoSs). The measurements of mass, radius, and recent detections of gravitational waves (GWs) have been essential in constraining the EoSs. Apart from direct observational signatures, physicists have also relied on indirect methods to draw inferences from NS properties. These include implementing Bayesian statistics \citep{Patra:2022yqc, Roy:2024sjx, Beznogov:2024vcv, Char:2020utj, Tewari:2024qit}, machine learning \citep{Chatterjee:2023ecc, Brandes:2024vhw, Carvalho:2023ele, Ventagli:2024xsh}, or studying universal relations (URs) \citep{Konstantinou_2022, niko1, Nath:2023gmu, laskos}. The former two techniques have been essential in drawing important conclusions regarding parameters, whereas the latter has proven to be vital for drawing relations among the parameters.

The initial step in studying URs involves developing EoS models. The EoSs can be motivated by a field theory approach or an agnostic approach. The former uses information from microphysics while including the interactions of particles to understand the behavior of matter at the core of NSs \citep{lattimer2004, tolos, oertel2015, weber1999, Annala:2019puf, HM, sabir}. Meanwhile, the agnostic approaches do not use any microphysics but make use of the current astrophysical constraints in developing the EoSs \citep{Most, greif, lind}. This approach provides more freedom to generate an ensemble of EoSs.

The physical properties that define a neutron star are expected to be influenced by the EoS. However, studies have demonstrated that certain combinations of these physical properties remain independent of the specific details of the EoS and adhere to universal relations \citep{YAGI20171}. Numerous studies have explored universal relations in neutron stars containing exotic matter \citep{luiz,lenka,Kumar:2023ojk}. Conventionally, URs have been studied in the context of measurable quantities like moment of inertia, mass quadrupole moment \citep{Staykov}, compactness \citep{Breu}, dimensionless tidal deformability \citep{Yagi_science}, and also in regards to the frequency of oscillating NSs \citep{Thakur:2024ijp, Guha:2024gfe, Roy:2023gzi, Ghosh:2024cay}. 

Astrophysical measurements play a crucial role in studying these URs. One such measurement is the gravitational redshift, defined as:
\begin{align}
Z_g=1/\sqrt{1-2GM/Rc^2}-1,
\end{align}
 where $G$ is the gravitational constant, $M$ is the mass of the star, $R$ is the radius of the star, and $c$ is the speed of light. Recent $Z_g$ measurements of RBS 1223, RX J0720.4-3125, and RX J1856.5-3754 with values of $0.16^{+0.03}_{-0.02}$, $0.205^{+0.006}_{-0.003}$, and $0.22^{+0.06}_{-0.12}$ of the $95\%$ highest posterior density respectively has proven to be a useful tool in constraining the EoSs of NSs \citep{Hambaryan_2014, Hambaryan}. However, the idea that $Z_g$ can be a useful tool in determining the properties of NSs is not new. Refs \citep{Buchdahl, Bondi, Hartle} showed that the surface redshift value for massive stars is $\le 2$. Further improvements \citep{Lindblom} bounded the $Z_g$ value for a $1.4 M_{\odot}$ star between $0.854 \geq Z_g \geq 0.184$. For the three redshift measurements  RBS 1223, RX J0720.4-3125, and RX J1856.5-3754, the isolated masses \citep{Tang_2020} as well as their bulk properties \citep{Luo_2022} have also been recently estimated. 

In this work, we try to analyze the URs of NSs in the context of the $Z_g$. So far in literature, only Ref \citep{Yang} has shown that there can exist a correlation between moment of inertia, gravitational redshift and gravitational binding energy. In this regard, it becomes necessary to analyze the URs of $Z_g$ with various parameters. For our analysis, we employ speed of sound ($c_s^{2}$) parameterization for construction of agnostic EoSs. Next we study the URs of $Z_g$ with regards to moment of inertia ($I$), dimensionless tidal deformability ($\bar{\lambda}$), mass quadrupole moment ($Q$), spin parameter ($\chi$) and average speed of sound ($<c_s^2>$). We also provide theoretical estimates of all these parameters using the URs for all three redshift measurements: RBS 1223, RX J0720.4-3125, and RX J1856.5-3754. From here on, we have used geometrized units (c = G = 1) for the rest of the article.


\section{Formalism} \label{sec:form}

\subsection{Construction of EoSs} \label{subsec:eos}
The slope of the EoSs is defined by the adiabatic speed of sound which can be denoted as $c_s$, where $c_s = \sqrt{ d p / d \epsilon}$. Here $p$ is the pressure and $\epsilon$ is the energy density. As $c_s$ is bounded between $0$ (from thermodynamic stability condition) and 1 (from the causality), hence it can be used to construct a family of EoSs in an agnostic manner by interpolating the EoS between the chiral effective field theory (EFT) and perturbative quantum chromodynamics (pQCD) limits. To begin with, we use a tabulated version of the Baym-Pethick-Sutherland (BPS model) for densities $n < 0.5 n_s$ \citep{Baym1971}. For densities ranging from $0.5 n_s \leq n \leq 1.1 n_s$ we use monotropes of $p(n) = k n^\Gamma$ where $\Gamma$ can range from $[1.77-3.23]$ and $k$ is obtained by matching it to BPS EoS \citep{Altiparmak_2022, Thakur:2024ijp}. During this process, it is ensured that pressure remains within the range defined in Ref \citep{Hebeler_2013}. For density ranges, $1.1\,n_s < n \leq 40\,n_s$, we use the sound-speed parametrization method introduced in \citep{Annala:2019puf, Altiparmak_2022, Thakur:2024ijp, ecker} defined as
\begin{align}
    n(\mu) = n_1 \exp\left({\int_{\mu_1}^{\mu} \dfrac{d\mu'}{\mu' c_s^{2}(\mu')}}\right)
\end{align}
where $n_1 = 1.1 n_s$ and $\mu_1 = \mu(n_1)$
The pressure can again be obtained from the number density as
\begin{align}
    p(\mu) = p_1 + \int_{\mu_1}^{\mu} d\mu' n(\mu')
\end{align}
where the constant $p_1$ is the pressure at $n_1$. To solve these two equations numerically we use a fixed number of segments between N(3,4,5,7) \citep{Altiparmak_2022} and use a piecewise linear interpolation as:
\begin{align}
    c_s^{2}(\mu) = \dfrac{(\mu_{i+1} - \mu) c_{s,i}^{2} + (\mu - \mu_i) c_{s,i+1}^{2}}{\mu_{i+1} - \mu_i}
\end{align}
where $\mu_i$ and $c_{s,i}^{2}$ being the chemical potential and the $c_s$ is randomly sampled between $\mu_1 \leq \mu_i \leq \mu_{N+1}$ and $0< c_{s,i}^{2} \leq 1$ at the $i$-th segment. Near the pQCD regime, we keep solutions whose pressure, density, and sound speed at $\mu_{i}=2.6\,{\rm GeV}$ are consistent with the parametrized perturbative result for cold quark matter in beta-equilibrium \citep{Kurkela_2010}. 

\begin{figure}
	\centering
	\begin{minipage}[ht]{0.5\textwidth} 
		\includegraphics[scale=0.57]{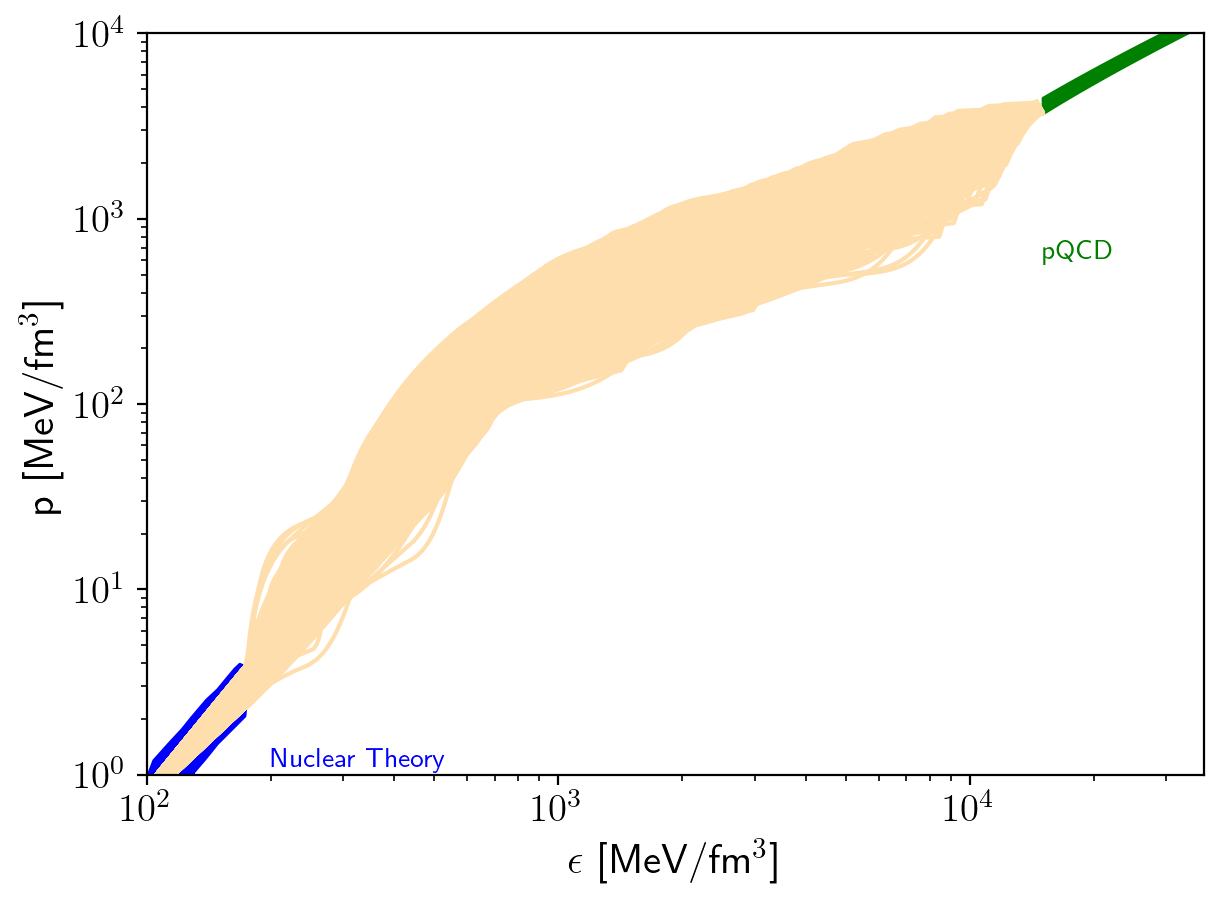}
	\end{minipage} 
	\begin{minipage}[ht]{0.5\textwidth} 
		\includegraphics[scale=0.60]{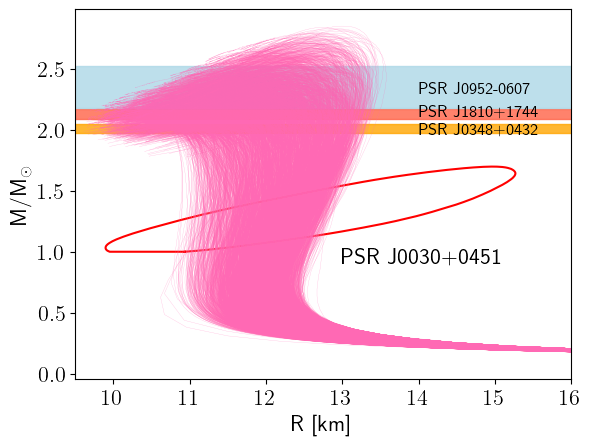}
	\end{minipage} 
	\caption{{\bf (Top):} \small The family of EoSs obtained using speed of sound parameterisation.{\bf (Bottom):} The MR sequences of the agnostic family of EoSs shown in pink along with various mass \citep{Antoniadis,Cromartie, Romani_2022} and radius \citep{Miller_2019} measurements of different pulsars.}
	\label{fig:EOS-MR}
\end{figure}

After obtaining the family of EoSs, we solve the Tolman-Oppenheimer-Volkoff equations \citep{TOV} to obtain the mass-radius (MR) relations. Following this we impose the following astrophysical constraints: 
\begin{itemize}
    \item The maximum mass for the EoSs was found to be $\geq 2M_{\odot}$ which comes from the mass measurements of PSR J0348+0432 \citep{Antoniadis} and PSR J0740+6620 \citep{Fonseca, Cromartie}.
    \item The EoSs were checked to satisfy the constraints imposed by the binary tidal deformability measurements from the low spin prior of GW 170817 $\tilde{\Lambda} \leq 720$ \citep{Abbott}. Where $\tilde{\Lambda}$ is given as:
    \begin{align}
        \tilde{\Lambda} = \dfrac{16}{13} \dfrac{(12 M_2 + M_1)M_1^4 \lambda_1 + (12 M_1 + M_2)M_2^4 \lambda_2}{(M_1 + M_2)^5}
    \end{align}
    with `$1,2$' denoting the two binary components and $\lambda_{1,2}$ their respective tidal deformability. The chrip mass  $\mathcal{M} = (M_1 M_2)^{3/5}(M_1 + M_2)^{-1/5} = 1.186 M_{\odot}$ for a mass ratio of $q = M_2/M_1 > 0.73$.
\end{itemize}
The EoSs and their corresponding MR curves obtained in this manner are shown in figure \ref{fig:EOS-MR}.

\subsection{Universal Relations} \label{subsec:UR}
The primary goal of URs is to explore the characteristics of NSs that are difficult to observe directly. Knowing one of the parameters helps us in estimating the other using the URs. It is widely recognized that universal relations are inherently linked to the correlation between various neutron star properties \citep{Pappas:2023PRD}, where a higher correlation between two quantities signify a greater universality. 


\section{Results}\label{sec:result}

We consider NS models that are described by their mass $M$, the magnitude of their spin angular momentum $J$ and angular frequency $\Omega$, its (spin-induced) quadrupole moment $Q$ and their moment of inertia $I \equiv J/\Omega$.  The dimensionless quantities are defined as $\bar{I} \equiv I/M^{3}$ and $\bar{Q} \equiv -Q/(M^{3} \chi^{2})$, where $\chi \equiv J/M^{2}$ is the dimensionless spin parameter. The dimensionless tidal deformability can be defined as $\bar{\lambda} = \lambda/M^5$, where $\lambda$ is the tidal deformability. The parameter $M \times \bar{f} / \chi$ where $\bar{f}$ = $\frac{\Omega}{2\times\pi}$. The average speed of sound is defined as $c_s^2=\partial p/\partial\epsilon$, integrated over the energy density \citep{saes1,saes2}:
\begin{equation}    
 \langle c_s^2\rangle \equiv \frac{1}{\epsilon_c}\int_0^{\epsilon_c} \dd\epsilon c_s^2(\epsilon)
\end{equation}
where $\epsilon_c$ is the central energy density of the star. The NSs are modelled using the {\ttfamily RNS code} \citep{stergioulas,nozawa} that helps us to determine the above quantities. We configured {\ttfamily RNS} with the finest grid which is $151$ $\times$ $301$ (angular $\times$ radial), and applied a tolerance of $10^{-4}$ for the specified parameter values. We make use of the agnostically generated EoSs and report the results below.

The relationship between $Z_g$ and other parameters of interest (except with dimensionless tidal deformability) can be fitted with the help of a logarithmic fitting function, which is shown as follows: 

\begin{align}\label{eq:fit}
    \log_{10}{y} = \Sigma_{i=0}^{4} a_{yi} \: \log_{10}(Z_g)^i
\end{align}

where $y=\bar Q$, $\bar I$, $M \times \bar{f} / \chi$, and $<c_s^2>$. The corresponding coefficients in the fits are listed in table \ref{fit-tab}. These fitting values are obtained through the use of smooth family of agnostic EoSs with $\Omega=480$Hz. Here, $Z_g$ is computed for a non-rotating star but with the same central density. It was found that the fitting was most accurate for the relation between $\bar{\lambda}$ and $Z_g$, upon including the exponential and linear terms in the fitting  eq \eqref{eq:fit} as below:
\begin{align}\label{eq:fit1}
    \log_{10}{y} = \Sigma_{i=0}^{4} l_i \: \log_{10}(Z_g)^i + l_5(Z_g) + l_6 (\exp{Z_g})
\end{align}

where $y=\bar \lambda$, and the coefficients are: $l_0$ = $97.33032$, $l_1$ = $241.45649$, $l_2$ = $209.56402$, $l_3$ = $89.10265$, $l_4$ = $15.47452$, $l_5$ = $-167.10804$, $l_6$ = $26.2434$. The fractional percentage error can be defined as $\lvert \Delta \rvert$ = $\lvert\frac{V_y - V_{fit} }{V_{fit}}\rvert$, where $V_y$ is the value of a parameter obtained from theoretical NS models and $V_{fit}$ is the value of the corresponding fitting function $\log_{10}y$. The value of both $V_y$ and $V_{fit}$ are corresponding to a specific value of $Z_g$.

\begin{table}
	\centering
	\begin{tabular}{@{\hspace{0.3cm}}c @{\hspace{0.3cm}}c @{\hspace{0.3cm}}c @{\hspace{0.3cm}}c @{\hspace{0.3cm}}c @{\hspace{0.3cm}}c}

		\hline
		\hline
		\noalign{\smallskip}
		   &{$a_{y0}$} &{$a_{y1}$}
		& {$a_{y2}$} & {$a_{y3}$} & {$a_{y4}$} \\
		\hline
		\noalign{\smallskip}
     $\frac{M\times\bar{f}}{\chi}$ & -1.4351 & 0.26991 & -0.61885 & 0.2966 & 0.27185\\
        $\bar{I}$ & 0.65139 & -0.10017 & 1.12242 & 0.34835 & 0.00771\\
        $\bar{Q}$ & -0.06672 & -0.6317 & 1.69581 & 1.54209 & 0.42514\\
       $<c_s^2>$ & 0.00902 & 2.00443 & 1.81783 & 1.51151 & 0.49134\\
		\noalign{\smallskip}
		\hline
		\hline
	\end{tabular}
	\caption{Fitting parameters for the universal relations.}
	\label{fit-tab}
\end{table}

\begin{figure}
    \centering
    \includegraphics[width=1.02\linewidth]{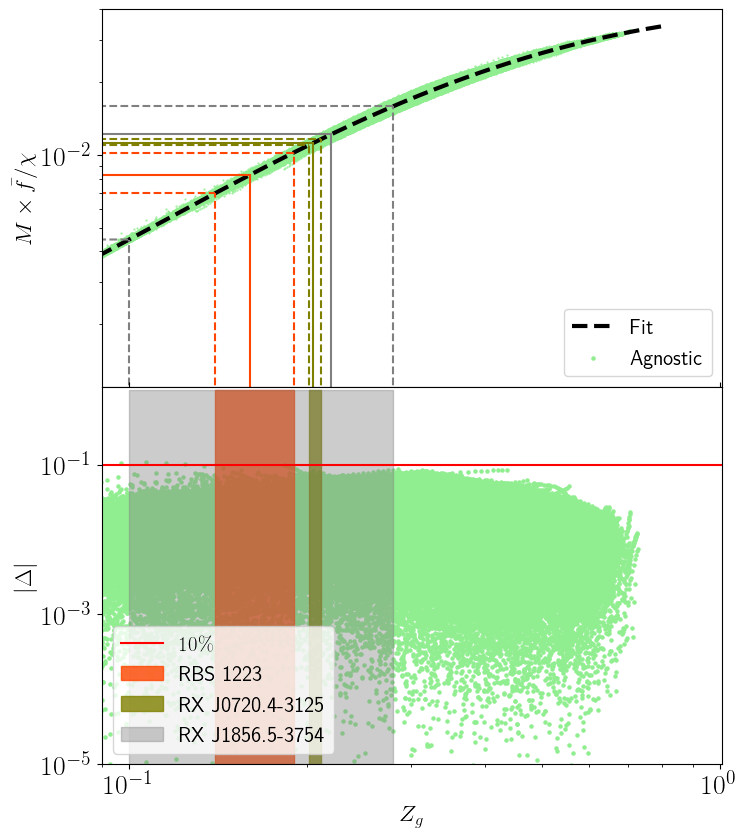}
    \caption{UR of $M \times \bar{f} / \chi$ with $Z_g$. The coefficients of the black dashed fitting curve is mentioned in the the table \ref{fit-tab}. The red horizontal line denotes 10\% tolerance limit. }
    \label{fig:Dzg}
\end{figure}

\begin{figure}
    \centering
    \includegraphics[width=1.02\linewidth]{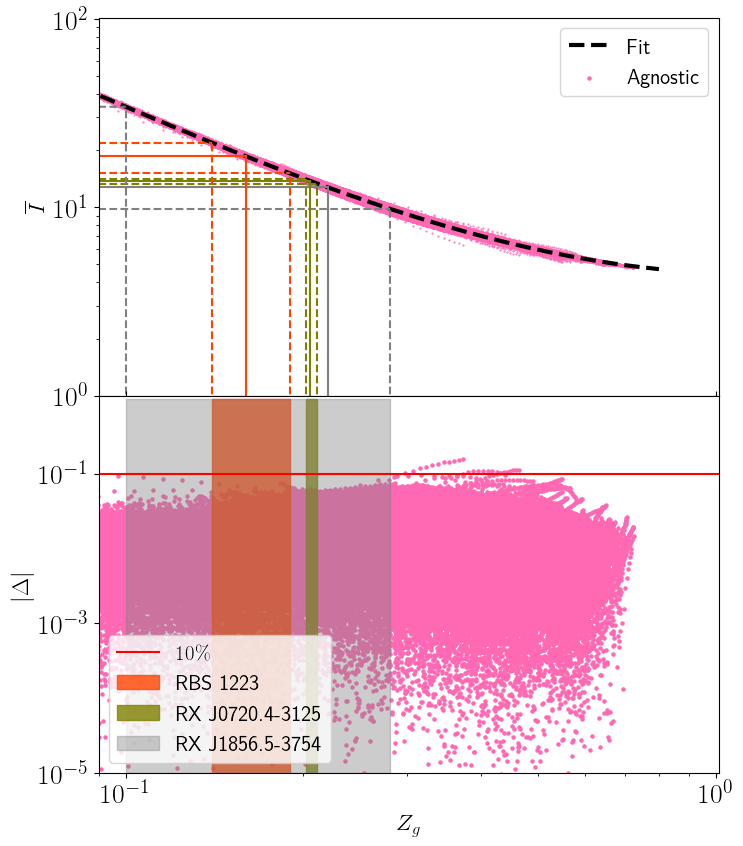}
    \caption{UR of $I$ with $Z_g$. The coefficients of the black dashed fitting curve is mentioned in the table \ref{fit-tab}. The red horizontal line denotes 10\% tolerance limit. }
    \label{fig:Izg}
\end{figure}

The values of fitting functions eq \eqref{eq:fit} and \eqref{eq:fit1} corresponding to the limiting values of uncertainty in observations provide us with the theoretical estimate of the various parameters of NS. 
In figure \ref{fig:Dzg} and \ref{fig:Izg}, the grey dashed lines depict the values of $M \times \bar{f} / \chi$ and $\bar{I}$ corresponding to the upper and lower limit of the observation J1856.5-3754 with the solid line representing their mean. The same observation is highlighted by the grey shade in the subfigure below. A similar nomenclature is followed for RBS 1223 (orange) and RX J0720.4-3125 (green). The red horizontal line in the bottom subfigure denotes a 10\% tolerance limit. This particular tolerance limit is chosen as agnostic EoSs following astrophysical constraints were found to satisfy this tolerance in I-love-Q UR analysis \citep{Nath:2023gmu}.

\begin{figure}
    \centering
    \includegraphics[width=1.02\linewidth]{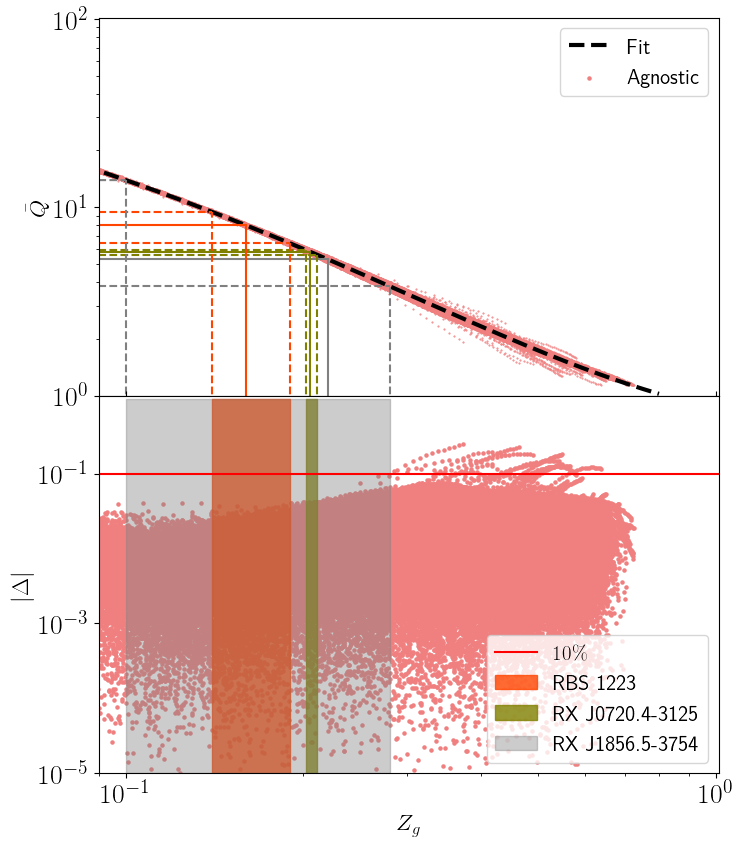}
    \caption{UR of $\bar{Q}$ with $Z_g$. The coefficients of the black dashed fitting curve is mentioned in the table \ref{fit-tab}. The red horizontal line denotes 10\% tolerance limit.}
    \label{fig:Qzg}
\end{figure}

Within the shaded region of the redshift observations, we find that the URs corresponding to $M \times \bar{f} / \chi$ and $\bar{I}$ follow the tolerance limit of 10\% as seen in the figures \ref{fig:Dzg} and \ref{fig:Izg} respectively. The relation between $\bar{Q}$ and $Z_g$ also follows the same as illustrated in the figure \ref{fig:Qzg}. However, the parameters $\bar{\lambda}$ and $<c_s^{2}>$ show a violation of the 10\% tolerance limit, thereby increasing the error in the theoretical estimates for these quantities as seen from table \ref{table_opt}.

The theoretical estimates shown in table \ref{table_opt} are estimated from the fitting functions of the URs. The redshift measurements RBS 1223, RX J0720.4-3125, and RX J1856.5-3754 are used to predict the parameter values using the fitting function. Corresponding to their mean and the standard deviation values, the functions in eq (\ref{eq:fit}), (\ref{eq:fit1}) and values in the table \ref{fit-tab} are used to obtain the values in the table \ref{table_opt}.

\begin{figure}
    \centering
    \includegraphics[width=1.02\linewidth]{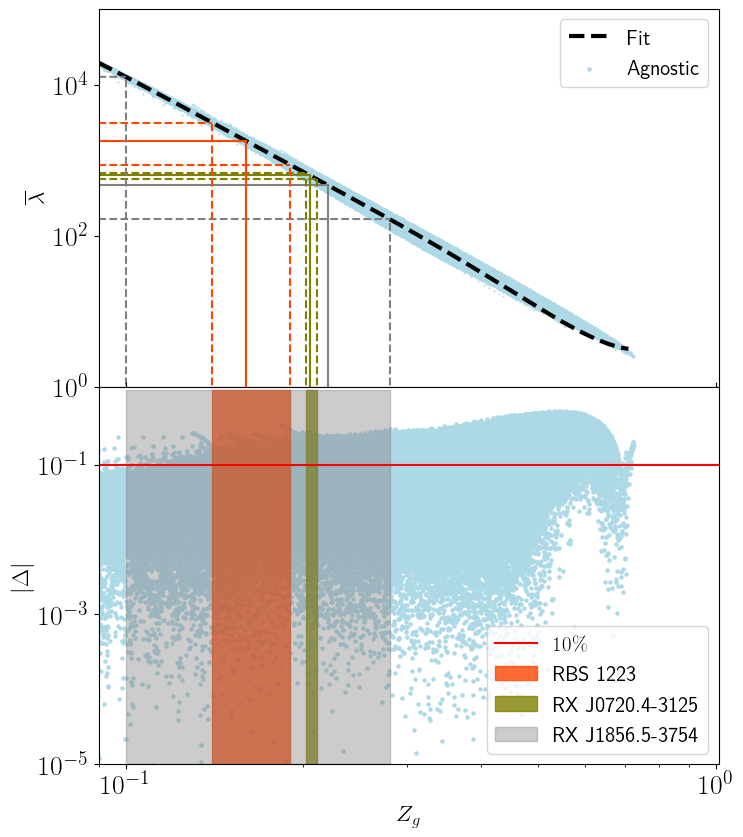}
    \caption{UR of $\bar \lambda$ with $Z_g$. The coefficients of the black dashed fitting curve is mentioned in the text. The the red horizontal line denotes 10\% tolerance limit.}
    \label{fig:lovezg}
\end{figure}

\begin{figure}
    \centering
    \includegraphics[width=1.02\linewidth]{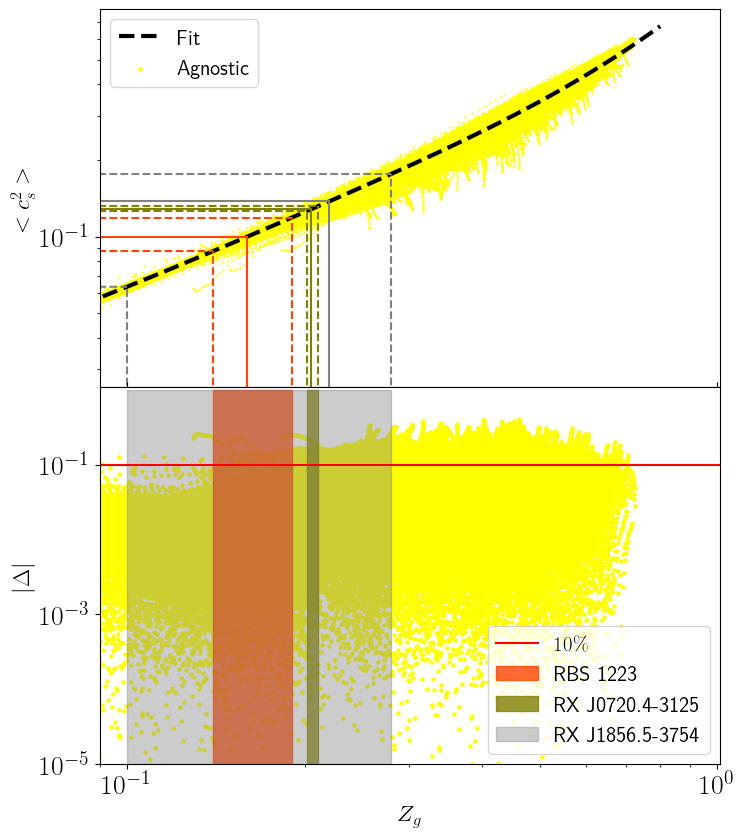}
    \caption{UR of $<c_s^2>$ with $Z_g$. The coefficients of the black dashed fitting curve is mentioned in the table \ref{fit-tab}. The red horizontal line denotes 10\% tolerance limit.}
    \label{fig:cs2zg}
\end{figure}

\begin{table*}
     \centering
      \caption{ Table showing the theoretical estimates of the parameters with respect to the URs drawn against $Z_g$.}
     \label{table_opt}
     \begin{tabular}{@{\hspace{0.31cm}}c @{\hspace{0.31cm}}c @{\hspace{0.31cm}}c @{\hspace{0.31cm}}c @{\hspace{0.31cm}}c @{\hspace{0.31cm}}c @{\hspace{0.31cm}}c@{\hspace{0.31cm}}c}
     \hline
     \hline \\
     
          & $\bar{I}$ & $\bar{\lambda}$&$\bar{Q}$ & $M \times \bar{f} / \chi$ & $<c_s^2>$ \\ \hline  \\
         RBS 1223 ($Z_g = 0.16^{+0.03}_{-0.02}$) &  $18.6^{-3.5}_{+3.4}$ & $1787^{-923}_{+1350}$ & $8.0^{-1.59}_{+1.44}$ & $0.008^{+0.002}_{-0.001}$ & $0.1^{+0.019}_{-0.012}$ \\  \\ \hline \\
         RX J0720.4-3125 ($Z_g = 0.205^{+0.006}_{-0.003}$)&  $13.8^{-0.4}_{+0.2}$ & $626^{-72}_{+40}$& $5.8^{-0.22}_{+0.11}$ & $0.011^{+0.0004}_{-0.0002}$ &$0.128^{+0.004}_{-0.002}$   \\ \\  \hline \\
          RX J1856.5-3754 ($Zg=0.22^{+0.06}_{-0.12}$) & $12.7^{-2.9}_{+21.4}$ & $464^{-299}_{+12194}$& $5.28^{-1.48}_{+8.64}$  & $0.012^{+0.004}_{-0.008}$& $0.138^{+0.04}_{-0.07}$ \\ \\  \hline \\
        
           \\ \hline\hline \\
     \end{tabular}
    
 \end{table*}

\begin{figure}
    \centering
    \includegraphics[width=0.9\linewidth]{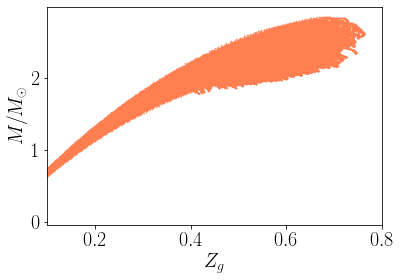}
    \caption{Variation of $Z_g$ with mass of the stars.}
    \label{fig:massZg}
\end{figure}
As the EoSs shown in the figure \ref{fig:EOS-MR} follow the astrophysical constraints, they can provide us with an upper limit on the value of $Z_g$. In the figure \ref{fig:massZg}, we plot the variation of the mass with the gravitational redshift and find that the maximum value of gravitational redshift attained is: $Z_{g}(max) \leq 0.763$ which further constraints the previous maximum estimates of $\le 2$ \citep{Buchdahl, Bondi, Hartle}. The range of values of $Z_g$ for a $1.4 M_{\odot}$ NS can also be seen to agree with the limiting values for the same provided by the Ref \citep{Lindblom}, i.e. $0.854 \geq Z_g \geq 0.184$.

\section{Summary and Conclusion} \label{sec:Summ}
In this work, we have examined the correlation between various properties of NSs and the gravitational redshift using URs. Our findings indicate that the quantities $M \times \bar{f}/ \chi$, $\bar{I}$, and $\bar{Q}$ are quasi-universal, with most of the observations lying below the 10\% tolerance limit. We further show that $\bar{\lambda}$ and $<c_s^{2}>$ tend to violate the 10\% tolerance limit in universality.

Knowledge about the UR of $M\times \bar{f}/ \chi$ can help us estimate the rotational properties if the value of $Z_g$ is also known. In the event of an observation of redshift from sources like NS, the ratio $\bar{f}/{\chi}$ can be determined within a certain margin of error, provided the mass of the star is known. For the UR with $\bar{I}$ and $\bar{Q}$ within the observational region of redshifts, the error tolerance lies within the 10\% limit. The scenario changes when we analyze the URs with respect to $\bar{\lambda}$ and $<c_s^2>$. For these two particular parameters, we see a higher deviation in the fitting function, resulting in a violation of the 10\% limit. Traditionally, we have seen that the $\bar{I}$, $\bar{\lambda}$, and $\bar{Q}$ show higher universality among themselves (the I-love-Q relations), suggesting that if there exists a universal relation between any two parameters then the third parameter will also be universal. Contrary to previous studies, we see that though $Z_g$ is universal with $\bar{I}$ and $\bar{Q}$, it is not universal with $\bar{\lambda}$. From the $<c_s^2>$-$Z_g$ analysis we see that lower values of $<c_s^2>$ is found for smaller $Z_g$ values.

We then utilize the URs and redshift measurements of RBS 1223, RX J0720.4-3125, and RX J1856.5-3754 to offer a theoretical estimate of these parameters using the fitting functions. The theoretical estimates calculated using the fitting functions of URs provide a scope to recover the values of stellar properties that are not directly observable. However, the accuracy of the estimates might vary depending on how universal the relations are. The greater the correlation of a parameter with $Z_g$, the higher the accuracy of the theoretical estimate. 

Depending on the universality of a relation, other parameters $\bar Q$, $\bar I$ and $\bar \lambda$ can be depicted with a certain degree of accuracy. For each of the three gravitational redshift measurements, the probable range of each parameter is shown in \ref{table_opt}. Luo et al. \citep{Luo_2022}, using a Bayesian approach, provided mass and radius estimates for three measurements along with 68\% confidence intervals for the tidal Love number. While our method differs from theirs, a comparison reveals that our theoretical estimates are in close agreement with their results. As shown in the figure \ref{table_opt2}, it is important to note that their results are based on a 68\% confidence interval, while our theoretical estimates are derived from a 95\% confidence interval based on redshift measurements. We see the $\bar{\lambda}$ estimate of RX J0720.4-3125 closely agrees with the Bayesian estimate. The estimates from RBS 1223 are also well within the estimated standard deviation from their work. The results from RX J1856.5-3754 show large deviations. This is attributed to the fact that this observation is accompanied by significant indeterminacy and the relation of $\bar{\lambda}$ with $Z_g$ has lower universality, as seen in the deviation plot \ref{fig:lovezg}. Since the uncertainty in the measured redshift for RX J0720.4-3125 is the least, the theoretical estimate closely agrees with the Bayesian estimate. Our analysis provides a new window to look into URs with regards to $Z_g$ .We have also demonstrated that, for stars consistent with current astrophysical constraints, the upper limit of $Z_g$ is $\sim$ $0.763$. It shows that theoretical estimates are accurate for observations with less uncertainty and can be used as an alternative for statistical analysis. The entire method outlined in this article may not be the most sophisticated or precise; however, it is certainly one of the feasible ways to extract values of NS properties from redshift observations. It is to be noted that $Z_g$ considered in our work is the surface redshift for a static star. In \ref{app:compare} we show how our results can change quantitatively upon considering the redshift of the rotating star. We also show that the qualitative analysis will hold true along with the URs.
\begin{table}
     \centering
      \caption{ Comparison of $\bar{\lambda}$ estimates of our work with Luo et al. \citep{Luo_2022} (for only 4P model with 68\% confidence interval).}
     \label{table_opt2}
     \begin{tabular}{@{\hspace{1.0cm}}c @{\hspace{1.0cm}}c @{\hspace{1.0cm}}c}
     \hline
     \hline \\
     
          & Luo et al.& This work   \\ \hline  \\
         RBS 1223 & $420^{+3260}_{-370}$ &  $1787^{+1350}_{-923}$  \\  \\ \hline \\
         RX J0720.4-3125 &  $641^{+56}_{-48}$ & $626^{+40}_{-72}$  \\ \\  \hline \\
          RX J1856.5-3754 & $1460^{+890}_{-980}$ & $464^{+12194}_{-299}$ \\ \\  \hline \\
        
           \\ \hline\hline \\
     \end{tabular}
    
 \end{table}
\section*{Acknowledgement}
SC would like to acknowledge the Prime Minister's Research Fellowship (PMRF), Ministry of Education Govt. of India, for a graduate fellowship. KKN would like to acknowledge the Department of Atomic Energy (DAE), Govt. of India, for sponsoring the fellowship covered under the sub-project no. RIN4001-SPS (Basic research in Physical Sciences). The authors would like to thank Deeptak Biswas and Mahammad Sabir Ali for careful reading of the manuscript.

\bibliography{main}





\appendix
\section{Effects of rotation on redshifts}
\label{app:compare}
\begin{figure}
    \centering
    \includegraphics[width=1.02\linewidth]{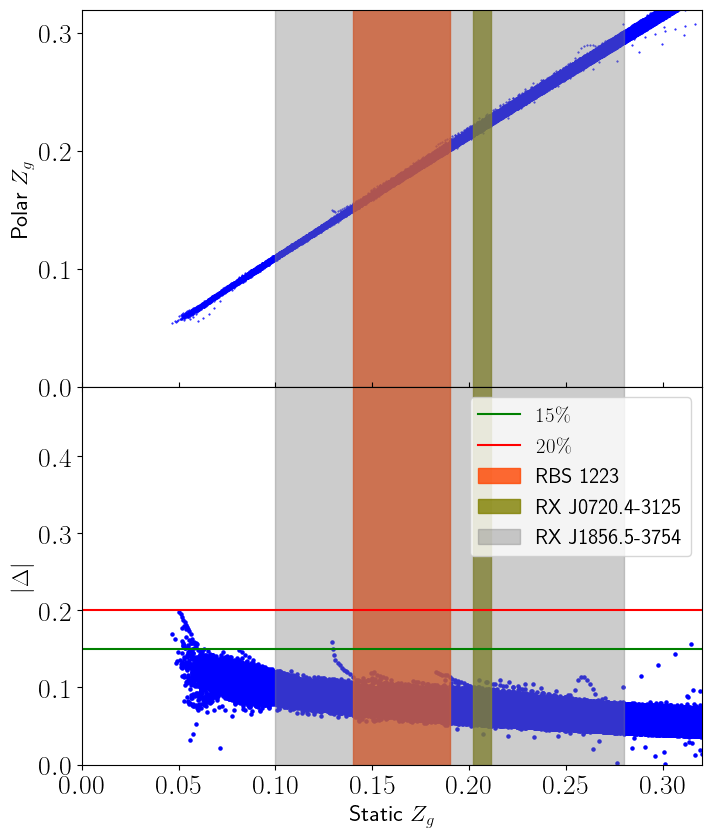}
    \caption{The upper panel shows the variation of the polar redshift with the static redshift. The lower panel represents the fractional relative difference between them.}
    \label{Zg-compare}
\end{figure}

The calculations explored in this study consider the ideal scenario where the surface redshift of static non-rotating stars was used for our calculations. In this section, we calculate the variation of polar redshift (for the rotating stars) with respect to the static redshift for all our EoSs as shown in figure \ref{Zg-compare}. We set up the rotating NS code RNS \citep{stergioulas,nozawa} (highest grid: DMDIV = 151 and DSDIV = 301 ) with a rotating frequency of $\sim 480$ $Hz$. We show the fractional relative difference between the two redshifts ($|\Delta| = |{\frac{\text{Polar} \: Z_g - \text{Static} \: Z_g}{\text{Static} \: Z_g}}|$) in the lower panel of the same figure. As per the figure, the fractional relative difference between them never exceeds $20\%$ and is within the range of our observed redshift measurements. We always find them to be less than $15\%$ (for most EoSs, the difference is less than $10\%$). This shows that if one considers the redshift measurements of the rotating star to compute the universal relations, the qualitative analysis will remain the same, only the coefficients of the fitting parameter may vary.

\end{document}